# A Uniform Quantum Computing Model Based on Virtual Quantum Processors

Georg Gesek
CTO Research & Development

QMware AG
St. Gallen, Switzerland
georg.gesek@qm-ware.com
ORCID:0000-0002-2967-1559

*Abstract*— Quantum Computers, once fully realized, can represent an exponential boost in computing power for certain applications and would likely serve alongside other classical devices in data centers of the future. However, the computational power of the current quantum computing era, referred to as Noisy Intermediate-Scale Quantum, or NISQ, is severely limited because of environmental and intrinsic noise, as well as the very low connectivity between the qubits, compared to their total amount within the current devices. Even in quantum computers with a large number of qubits, most of those qubits are assigned to error correction, leaving only a few, so-called Algorithmic Qubits, to perform actual calculations. Experts have estimated that the NISQ era could last years, if not decades. The uncertainty surrounding the development of fault-tolerant machines and cloud-based services, in itself, has led to a reluctance among many organizations to invest heavily in quantum software development. Industry leaders worry that a new technology or new device could make software developed for a competing system obsolete and, as a result, squander the investment. This paper argues that quantum simulators provide a feasible, cost-effective transition to the quantum era by allowing developers the ability to run quantum programs in a high-performance computing environment that is hardware agnostic, while still anticipating the unique computational abilities of future quantum systems. We propose that a virtual quantum processor that emulates a generic hybrid quantum machine can serve as a logical version of quantum computing hardware that is also capable of being run on today's machines. This hybrid quantum machine powers quantum-logical computations, which are substitutable by future native quantum processing units (QPU) to improve the design and execution of quantum algorithms, and provide a training space for high-performance computing centers during and after this transition into a fault-tolerant and scalable quantum era.

## I. INTRODUCTION

In order to determine how current Turing machines could emulate a Logical Quantum Processor, which is a generic representation of any physical implementation of such a technology, one has to scrutinize the fundamentals of Turing and quantum machines. First of all, we can identify quantum computing as a special application of quantum physics. Because physicists established a very distinct mathematical model of quantum physics in the 20th century, which is based on linear algebra in multidimensional complex vector spaces, so-called Hilbert spaces, we know that quantum physics can be calculated within classical computers. This is done by calculating matrices of floating-point representations, such as the IEEE754 binary floating-point format, in software that is called a "Quantum Computing Simulator". The term simulation refers to the representation of so-called 'Bloch spheres' |1| in Turing machines. Thus, a classical computer is able to reproduce the calculations that a quantum computer makes on the state vectors of its qubits accounting for the qubit's connections representing quantum logic gates, building the so-called quantum circuit without errors. The difference between classical and quantum computers and the reason scientists are eager to implement and use quantum computers, is the native processing of quantum information, which scales exponentially better while calculating these large matrices with entangled state vectors. If quantum computers only relied on the superposition of quantum information, meaning single qubit operations, they are in fact being executed in our classical computers nearly as efficient[1] as within quantum computers. But due to the possible entanglement of state vectors, the respective matrices calculations scale computer time exponentially within a Turing machine, while they don't in a quantum machine.

## II. THE MAGIC OF ENTANGLEMENT

The human brain, as a product of evolution in our macroscopic world, is specialized to perceive information from the senses in order to match them with previous impressions and comprehend them with preinstalled or learned algorithms, which lead to models of thinking and understanding about the world around us that are referred to as general intelligence. But, in fact, the so-called generality of our mind holds us back from the underlying realm of conscious reality, the world of quantum physics. Just as the famous quote of Richard Feynman suggests "I think I can safely say that nobody really understands quantum mechanics.", human perception is puzzled by the strange behavior of quantum systems, which could not be unraveled, at least during the 20th century.

---

[1] The matrices, representing the state vectors of the qubits, are operated with a certain set of classical operations for any superposition within a Turing machine, which means a shift of the two angles of the state vector. In a quantum computer, this is called a rotation gate on a single qubit and can be ideally carried out, in a single step, in parallel for any number of qubits. The Turing machine can also parallelize the calculation of superposition with such a linear extension of calculation power. Thus, there is no significant advantage for the quantum machine.

Computer engineers must understand the difference between classical and quantum information in order to create a virtual quantum processor, which is desirable to be implemented as a foundation of a hardware agnostic hybrid quantum computing model.

Therefore, let's begin with what we know about quantum entanglement, which, as mentioned, is one of the most important principles behind computer speed-up. The simplest system of a quantum entangled state is a two-particle system with just two plumbable qualities for each particle, which translates into a two-qubit-system. With the physical notation of Bra and Ket vectors, and the two possible outcomes of a measurement such as 0 & 1, we can write the state of the entangled system S:

$$|S\rangle = \alpha|00\rangle + \beta|11\rangle \qquad (1)$$

Where $\alpha$, $\beta$ are normalized complex numbers, so that the sum of their complex conjugate squares equals the unit:

$$\alpha\alpha^* + \beta\beta^* = 1 \qquad (2)$$

This is the quantum mechanical formalism to express what happens when a measurement of the State |S> takes place. The result can only be what the measurement apparatus allows to perceive, not what the quantum state "in reality" is. We have to take this seriously: our concept of reality implies what is measured, not the previous quantum state, which can be in a superposition of possible measurement results or can be entangled with other quantum systems. This definition of reality is due to the brain as an evolutionary product, which uses senses to perceive what is measured. In fact, to measure the outcome of the quantum state, the measurement apparatus must be entangled with the quantum system in question. This is also the reason a real measurement cannot be simulated with a deterministic Turing machine (DTM), because the outcome is purely probabilistic. But we can easily imagine connecting some physical (quantum) system outside of the DTM in order to retrieve a non-causal input regarding the computations at any given stage of its program. After such a measurement is simulated, we can then decide the probabilistic outcome of the state vector measurement with a true random input, which extends the capabilities of our DTM to a non-deterministic one. The normalization of the coefficients $\alpha$ and $\beta$ in (2) makes them available for the so-called Copenhagen interpretation of quantum mechanics, for their complex conjugate squares are interpreted as the probabilities for the possible outcomes, if one measures exactly for the given states |00> and |11>[2].

$$P(00) = \alpha\alpha^* \qquad P(11) = \beta\beta^*$$
$$P(01) = 0 \qquad P(10) = 0 \qquad (3)$$

In the language of information theory, we ask the quantum system which of these states will it present to the measurement apparatus. The quantum system's answer can then be derived from the interaction between itself and the measurement apparatus. In this (inter)action, both of them will be forced into new quantum states, meaning, both of them will be changed simultaneously.

Now, the state |S> in (1) is an entangled one, since it is constituted by two quantum systems, which can be both in the states |0> and |1>. In the non-entangled case, this would correspond to four possible outcomes of the two independent quantum systems, namely |00>, |10>, |11> and |01>. But since the quantum systems are represented by the state |S>, they are fully entangled. This means, if one of the two systems are measured, the other system's future is determined. For any measurement that takes place after that, the answer will be the same as the previously measured quantum system, as long as the same question has been asked, which means the same quality of the system has been measured.

The funny thing for a human's perception is that this quantum mechanism of entanglement functions over arbitrary distances in no time. But this does not mean that quantum entanglement is weird or spooky, as Albert Einstein stated, it just reminds us of the brain's functional principles that only retrieves information from measurements in space-time. This type of information we refer to as classical, while the information within the state vector of a quantum system, or a qubit, we recognize as quantum information.

### III. (Quantum) Information

We learn from the observations of quantum systems that there is a profound dependency between classical and quantum information. Thus, the question for the best technical implementation of hybrid quantum computers should be derived from the exact understanding of the dependencies between a bit of classical information **b** and a bit of quantum information |**b**>.

In order to retrieve **b**, one has to measure |**b**>. In our so-called classical world, we perceive only **b**. But, in fact, everything and everyone in this universe are quantum systems, regardless of size. The only difference between the 'macroscopic' and the 'microscopic' world, which is again a distinction based on purely subjective size of the body, is simply the number of quantum measurements per unit time, or particle interactions in terms of classical physics. Therefore, we are used to a myriad of quantum measurements automatically carried out within our macroscopic bodies, which give us the image of a smooth, or analogue -- but ultimately illusory -- reality of a world consisting of arbitrary amounts of features.

The same is true for computers. If it wasn't for ongoing, self-acting interactions between elementary particles in integrated circuits, the quantum computer would need to be invented first, in order to retrieve the classical information that is being calculated. But integration efforts already led to the point in which interactions between the electrons in integrated circuits and the crystal lattice become less probable, leaving time for quantum effects, such as the unwanted tunneling of electrons through the insulated gate of a MOSFET.

---

[2] Although it seems just to be a little formality, the difference between the notation of the quantum state in brackets, such as |00> and its corresponding measurement outcome 00 is enormous, since it distinguishes the quantum space with all its features like superposition, entanglement, non-locality and non-causality from our classical, local and causal reality. Thus, we should be aware and attentive of this important differentiation in our notation as well as notion. Between |00> and 00 resides the measurement process, which is not only by technical means, but also conceptual a big step, elevating our system qualities from the quantum informational plane to the meta-informational, classical layer.

As we can see now, classical correlates to quantum information as meta information since classical information would not exist without quantum information. The other way around, however, sees no such restriction. Apart from this distinction, classical information functions as a subset of quantum information. Since in computer and data science, we are very familiar with the concept of meta information, now it becomes much clearer how we should deal with classical and quantum information within computational models. Like the instruction set of a classical processor represents the meta information to the data processed, classical information represents the meta information on quantum information processed. There is no need to separately process them, if the processor has the capability of both classical and qubit registers. This insight has a fundamental influence to future quantum processor developments, but is also valuable for the following discussion of a generalized hybrid quantum computing model.

The following figure [1] of the Bloch sphere shall guide our efforts with the unification of classical and quantum computation. The quantum states, represented by the vectors |Ψ>, are in two possible classical values after a measurement, namely 0 & 1 at the opposite directions of the z-axis.

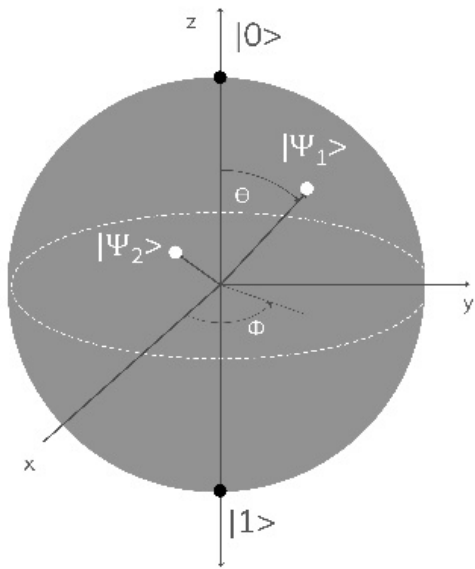

*Figure 1    Bloch Sphere*

It is important to note that these opposite directions represent orthonormal measurement outcomes because one can only measure 0 exclusive or 1. In fact, one could measure any axis through the origin of the sphere, but since this reflects a rotationally symmetric action, the given example is ubiquitous. Another important point raises a fundamental difference between the quantum states |0> & |1> of the state vector |Ψ> and the classical states 0 & 1 of a bit, after the measurement of the qubit. During a quantum computation, the state vector can reach any point on the Bloch sphere, without being measured. If it resides e.g., at |0> and we measure the qubit, we will get a classical 0 as an output. But if we don't measure the qubit, we won't get any classical information in space-time. For any other place on the sphere, where the angle θ differs from zero, the probability of a 0-measurement P(0) is:

$$P(0) = \Psi\Psi^* = \frac{1+\cos(\theta)}{2} \quad (4)$$

This probability is equal to the proportion of the area on the Bloch sphere below the circle of latitude, where |Ψ> points to, and its entire surface. In other words, the integral over all state vectors on the Bloch sphere equals to the unit 1. The physical meaning of this is simply that if you ask a quantum system about its state, you will always retrieve an answer. But there is a subtler feature with the probability equations (3) and (4): since a probability always means a ratio between some elements of a set and the whole set. Thus, the whole of a set of quantum information must be definite.

IV. THIRD QUANTIZATION

This feature of quantum information reflects the simple fact that it is quantized, but leads ultimately to an important technical implication, first described by Werner Heisenberg as the uncertainty principle:

$$\sigma_E \cdot \sigma_t \geq \frac{\hbar}{2} \quad (5)$$

The σ' stands for the standard deviations of energy (E) and time (t). As one can see, energy and time cannot become arbitrarily low because there is a small but finite limit set by half of the reduced Planck constant. Energy and time are a complementary pair, or in other words, canonically conjugated variables of physics. This unfolds during the measurement process: within the same measurement, one has to content oneself with a certain precision for the outcome, which is classical information. This is a fundamental principle of our universe, not an inadequacy in our measurement process, and couldn't be improved later. In other words, the number of distinguishable outcomes of measurements in any quantum system is finite, as long as the quantum system itself consists of a finite amount of energy and we only have finite time to measure it.

This concludes our observation from the probability equation for the qubit measurements. Furthermore, one could reason, we are only able to get a limited amount of classical information out of a quantum system due to the uncertainty principle, as the whole quantum system does not need to maintain more quantum states than necessary to provide us with this amount of information. Of course, it would shed more light on quantum information theory, if we would assume that the finite amount of quantum information, what every particle, molecule, planet, star or galaxy consists of, is the fundamental reason for the uncertainty principle and thus for the quantization of information in general.

If we follow this argument, we can calculate the number of quanta inhered by a quantum system in our observable universe. We refer from here to the concept of quanta as the fundamental property of quantum fields, as shown in quantum field theory |2|. Traditionally, the number of quanta has not been limited, but with our insights from quantum information theory, we can derive the limit for the information content, as a quantum system has to provide this, but not more, in order to give enough answers to our possible measurement procedures. Since this information is quantized – let the letter

I represent the number of quanta – we can calculate the minimum energy, which means I equals 1, corresponding to a system in our observable universe, since its wave function is bounded by our perspective as observers and the event horizon, within which we are able to make observations. Such a particle or quanta has its wavelength (λ) from us to the event horizon created by the Big Bang – or the starting point of the evolution of the universe, for any other cosmological theory apart from steady state – as follows:

$$\lambda = \frac{c}{H} \quad E = h \cdot \nu \cdot I \quad \nu = \frac{c}{\lambda} \quad \Rightarrow \quad E = h \cdot H \cdot I \quad (6)$$

c stands for the speed of light, H is the Hubble constant at the time of our measurement, and the formula for the Energy (E) is the usual one with the Planck constant h and the frequency of the wave function (ν), which translates in this particular case to H. For the ground state I equals one, and for the excited states I follows the natural numbers.

With (6) there is a definite connection between any object in the universe, regardless of its appearance, with a corresponding quantum field containing its information content, expressed by I and measured in the natural number of quanta. We may call this the Third Quantization, after the First Quantization initiated by Erwin Schrödinger and his wave equation and the Second Quantization introduced by Paul Dirac as the occupation number representation and extended by quantum field theory as the canonical quantization. The Third Quantization represents all physical values in quanta, even for space-time itself, since all physical objects represent a certain amount of energy.

Thus, even qubits are inherently quantized and the number of possible state vectors (I) is defined by

$$I = \frac{\Delta E}{h \cdot H} \quad with \quad \Delta E = \|E(|0\rangle) - E(|1\rangle)\| \quad (7)$$

Where ΔE represents the difference in energy between the two state vectors, which define the measurement values within the qubit and thus span the whole information space. It is noteworthy to realize that for such a physical qubit, or any quantum system, a Hilbert space transforms to its discrete equivalent with the Third Quantization.

Thus, any measurement along the z-axis forces the quantum system (qubit) into one of the quantum states |0> or |1>. But, contrary to mainstream literature of the 20th century, where the term of *collapsing wave functions* was coined, the qubit endures with its state vector and thus its wave function stays intact, but now is identified after the interaction with the measurement apparatus and, in fact both the qubit and the measurement mechanism remain entangled until one of the qubits interacts with another particle. So, if the qubit has been measured 0 and left undisturbed thereafter, the next measurement should also be a 0 value. For physical implementations of qubits, this can differ, but is then recognized as erroneous behavior of the qubit. The average time to this malfunction from the last 0-readout is called T1.

Such errors often occur with our current physical qubit implementations in native quantum registers, but they are not of interest in the following discussion about a uniform quantum computing model, where quantum registers are assumed to be free of errors. We ignore systematic as well as stochastic qubit imperfections at this stage, but are aware of the possibility of such unwanted influences from the rest of the universe into our quantum computer, which we start to describe in the following.

V. VIRTUAL QUANTUM PROCESSOR

The motivation to abstract from the physical implementation of a quantum processor is twofold:

First, we know that the limitations and erroneous behavior of native quantum registers will improve with advancements in engineering and manufacturing. They also are dependent on the topology of the qubits. While it is useful to emulate such physical systems, what quantum computer simulators can do to mitigate these errors in future applications is gratuitous in the field of high-performance gate-based quantum computing.

Another discipline of quantum computing, called 'analogue', has developed recently. Analogue quantum computing prepares a native quantum processor with all its errors as it is, in order to calculate a very special problem with this native Hamiltonian (the operator which describes the energy of a quantum system over time). However, this is not considered as a universal quantum computing system, as the gate-based variant that we are interested in, which uses quantum logic gates in a freely defined and variational array to process quantum information.

Second, we are looking to integrate all the functions of a quantum processor into a unified processor model, so we can emulate it and exchange any simulated parts of the model. For example, we could emulate quantum registers by appropriate physical implementations at any time and without changing any part of the whole computational stack above.

Both aspects let us concentrate on the development of a theoretically optimized high-performance universal quantum computing system, from which we can better anticipate the needed technological implications for its parts, compared to the opposite approach, which the industry is currently relying on to find suitable applications for the NISQ[3] systems.

Additionally, with this approach from scratch, we are able to use today's ready high-performance computing technology to implement this novel Unified Quantum Computing Model by its own means. The groundwork for this is the insight we have gained about the nature of classical information as meta information to quantum information and its unique act of being created during the measurement process. The insight of the Third Quantization lets us estimate the needed resolution of such a virtual qubit, in order to mimic the quantum system within a Turing machine.

According to (6) and (7), e.g., a RF qubit with a 1GHz range between its |0> and |1> state, inheres approximately 2 to the power of 90 possible quantum states; or in other words, with 90 classical bits, a Turing machine is capable of simulating the quantum bit (qubit) with no lack of accuracy. Of course, the actual readout accuracy of a physical qubit is many orders of magnitude lower and thus can be much easier emulated. On the other hand, a Turing simulated qubit with

---

[3] NISQ = Noisy Intermediate Scale Quantum Computer, which are state of the art right now

single (32 bit) or even double (64 bit) precision floating point representation has a much better resolution of the quantum information's intermediate representation than any of today's NISQ implementations.

The upshot is that we can store both an image of quantum and classical information within a classical memory!

But this alone does not give us a quantum computer. The latter is additionally capable of

A. *initializing qubits with classical meta information*

B. *initializing gate circuits between the qubits with classical meta information*

C. *processing the given quantum circuit by transforming all qubits by unitary matrix operations*

D. *measuring the qubits to retrieve classical information*

E. *processing classical information*

Literature sometimes refers to a quantum Turing machine as a deterministic Turing machine transformed by the simple exchange of the Turing machine's classical, discrete bit space by a Hilbert space. Now, we see that this isn't sufficient. The first novelty simplifies the simulation of qubit registers by the merging of an image of quantum and classical information within a classical memory. Of course, classically stored quantum information cannot be processed by a native QPU, but indeed by a virtual one.

In order to leverage the exponential advantage for certain quantum algorithms, a native QPU has to be integrated into such a system. Thus, the technical solution we are looking for is the hereby described Uniform Quantum Computing Model, which lets the applications run with both inbuilt virtual and native quantum processors, in parallel.

The second inaccuracy about the common description of a Quantum Turing machine is the important indeterminism for the measurement procedure. This is not difficult to implement into high-performance deterministic Turing machines because a simple sensor readout within a chassis with fluctuating bits of the length of the qubit accuracy would provide adequate randomness for our measurement simulation.

With a native QPU, this comes for free, of course. But we insert it into our model for a Universal Quantum machine as shown in [2], in order to be accurate. The precision of the full functionality of a native quantum computer, as well as the information theoretical complexity of its components, lets us replace any such part of the native quantum processor by an adequate simulation, optimized in performance for, what we can call **Advanced Quantum Inspired Computing (AQIC)**.

The third feature of native quantum computers, which is not represented within the conventional Turing model, is the measurement process. There, we must incorporate the uniqueness of quantum information in the universe. This means, if we measure a native qubit, we influence -- entangle -- with its stored quantum information to perceive its meta representation into our measurement apparatus as classical information. This is the reason why native quantum information cannot be copied, like its classical counterpart. Any such process of copying inheres a readout, which changes the source instantaneously. But of course, our classical image of the quantum information can be copied. We are even able to initialize a qubit, according to the precision of the apparatus, with very high accuracy as a physical representation of our quantum information image within the classical memory.

Under these aspects, it is even possible to create a close-to copying mechanism for native quantum information, where the accuracy of the copying process is higher than the gate fidelity, and thus functions as a good reasonable approximation for a certain algorithm.

Based on the claims *A.* to *E.* [2] shows the resulting block diagram of such a Universal Quantum machine. One can imagine the parts separate from the read-write tape [2A] to be the actual quantum processor. With [2A] it is also possible to process classical information, thus then considered as a Hybrid Quantum Processor.

According to the concepts and relations between quantum and classical information, displayed in chapter III, it is manifest that classical algorithms are a subset of quantum ones. In a technological sense, one could use qubits of a quantum register just with their |0⟩ |1⟩ values and thus representing the duality of a classical- with a quantum bit.

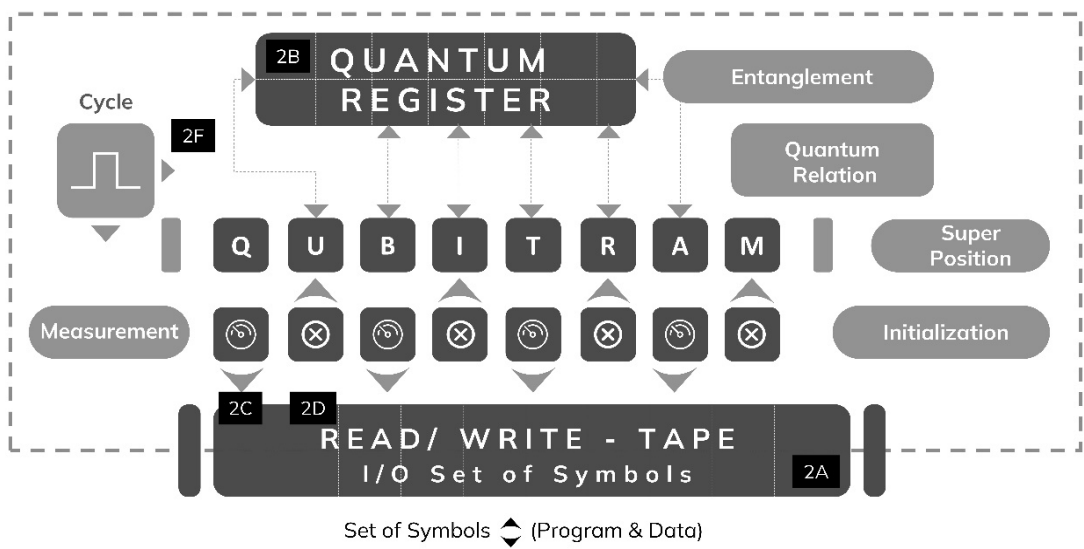

*Figure 2 Universal Quantum Machine*

For that reason [2] depicts, in fact, a Universal Quantum machine, since it provides a classical memory [2A] as well as quantum memory [2E], which both can be technically implemented as random access memories (RAM).

As required in {A}, [2D] initializes the random-access qubit memory (QRAM) [2E] by creating certain state vectors in its respective Bloch spheres, controlled by classical meta information, which represent the two degrees of freedom for each qubit, namely the two angles θ and Φ, defining the point where the state vector touches the surface of the Bloch sphere.

The requirement in {B} is represented by the relation between the QRAM [2E] and the quantum register [2B]. The processing {C} of the quantum gates within the quantum register is timed with the cycle generator [2F]. The measurement apparatus [2C] synchronously measures {D} the qubits after the processing of the quantum circuit and stores the classical values into the RAM [2A], which concludes the quantum part of the calculation handing it over to the classical post processing {E}.

Within the technical implementation of such a quantum processing unit (QPU), which is shown in [3], we anticipate the qubits [3E] and the quantum register [3B] within one physical system, shown as the quantum gates. This is owed to the fact, that our current quantum technological engineering skills are not yet sufficient to build a reliable QRAM.

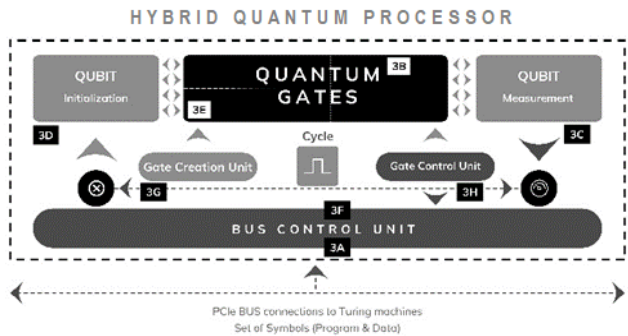

*Figure 3 Technical implementation of a QPU*

Furthermore, the gate creation [3G] and gate control [3H] units are certain hardware implementations, depending on the physical structure of the qubits, such as LASER or microwave pulse generators.

The bus control unit [3A], represents the classical cache memory and logic unit, which performs pre- and- post processing to the quantum circuits and connects the hybrid quantum processor to the external Turing machines. These handle large data transfers and storage needs to complement high-performance computing, as quantum computers are the highest performing information processing machines we can imagine. In order to technically accomplish this goal, we have to merge classical high-performance clusters (HPC) with QPUs during the next crucial step.

## VI. Uniform Information Processing

As previously mentioned, we are currently in a technological state where we haven't yet managed to implement a high-quality, highspeed quantum processing unit. We are still in the noisy intermediate scale quantum computing era (NISQ). This is obvious as we know the simulations of quantum computers on our classical Turing machines still perform much better for practical use cases, rather than on any physical quantum computing equipment.

On the other hand, we can only speculate as to the approach or vendor that will prevail with the task of high-performance quantum computing. The only thing guaranteed is the significant progress made over the last couple of years, which is likely to accelerate, given the large amounts of private and public money being poured into quantum technologies right now, with more accomplishments expected within the coming years.

Regardless of the certain construction of future technologies, we have already laid out the relevant features of such a computing system within the previous chapters. Since our Turing machines are already so powerful, we can make use of both the vast transactional calculation performance of our HPCs and the inherent quantum information logic.

The proposition here is the **Uniform Quantum Computing Model**. This is the idea to commonly store classical information alongside the classical representation of quantum information within **Bloch registers (BREGs)** and to compute these BREGs with a **Virtual Quantum Processing Unit (vQPU)**.

A vQPU consists of the same technological components as shown in [3], but implemented as software code within a large main memory of a Turing machine as depicted in [4]. In this uniform computing architecture, the main memory [4A] is accessible by all of the different processing units [4B] in the same manner and to its whole extent. Each of the processing units (CPU = Central Processing Unit; GPU = Graphical Processing Unit; NMPU = NeuroMorphic Processing Unit; GQPU = Gate based Quantum Processing Unit; QAPU = Quantum Annealing Processing Unit; DPU = Data Processing Unit) is either a physical implementation or a virtual processor. Within this paper, we analyze the gate-based quantum processing unit as a virtual processor and assume all the others as physical implementations. But one can follow the same logic for any other processing unit to be implemented as a virtual instance.

The minimum physical requirements are the central main memory [4A], at least one physical processing unit [4B] – at the moment the most advanced ones are the CISC, RISC and graphics processing architectures – the data processing unit [4D] with its bridge functionality between the internal and external systems, and the memory bus systems [4E] between the physical processing units [4B] and the main memory [4A]. For performance reasons, the cache coherency interconnect (CXL) [4C] should also be physical, but this is not mandatory.

With this architecture, one can implement the Uniform Computing Model in general, as well as the one for Hybrid Quantum Computing, in a very efficient and high-performance manner. All different types of processing units and, furthermore, any possible type of quantum processing unit - such as gate-based or annealing systems - are capable of being integrated. The physical type of the qubit registers is irrelevant for its functionality, only performance and quality constraints will be passed through the computational stack, depicted in [5].

This is possible due to the previously generic virtualization of the whole functionality of the quantum processors for any of the functions *A.* to *E.* in chapter V, which translates into a Turing machine's instruction set as follows:

Bloch sphere, to e.g., $2^{16}$ as a sufficient number. Such optimization parameters in the memory representation of quantum information have to be provided as meta information from the application layer [5E] via the kernel scheduler API [5D] to the inner core of the Hybrid Quantum Operating System [5G] in order to efficiently use the overall transactional computation power of the Uniform Quantum Computing Model.

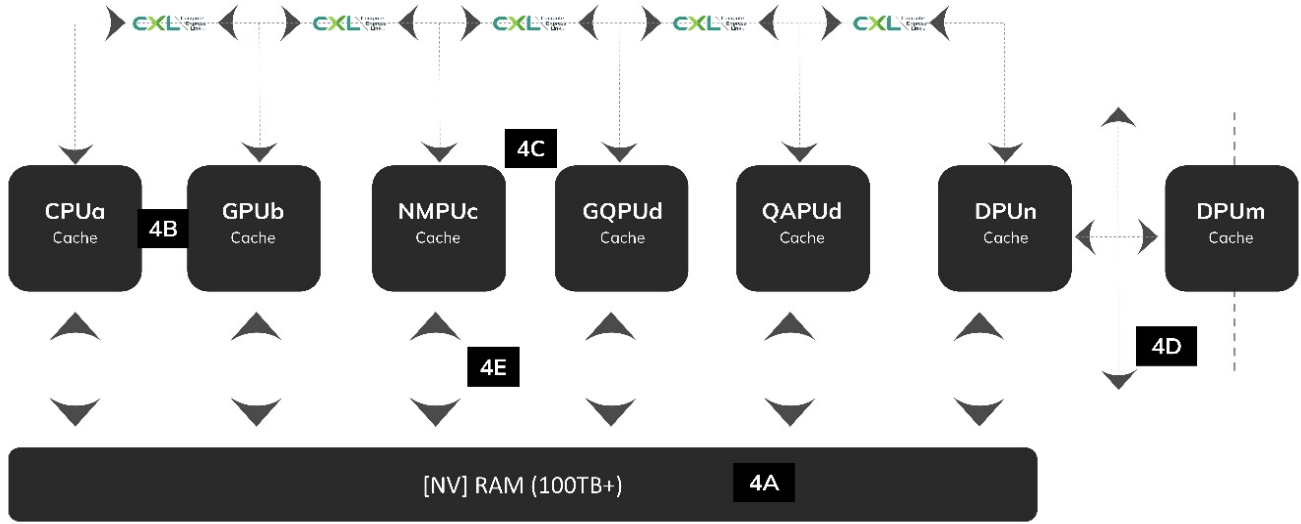

*Figure 4 Hybrid Processing Memory Centric Computing Architecture*

*A. initializing qubits with classical meta information*

This task requires the memory pattern translation [5B] to Bloch registers.

Of course, our current quantum circuit simulators just allocate the main memory for the storage of the linear matrices which are later computed with exponential time, in case entanglement occurs. But this is not an optimized process regarding the specifics of a certain quantum circuit. For example, the resolution depth of a qubit could be reduced from double precision, which means $2^{64}$ distinguishable points on a

*B. initializing gate circuits between the qubits with clasical meta information*

The same kernel scheduler API [5D] is used to transfer the required classical meta information for the quantum gate circuit to the memory pattern translation layer [5B]. In the case of physical qubit registers, this information is computed by the gate control unit [3H] of the native quantum processor. If the quantum processor is to be virtual, the gate matrices are constructed with this information within the main memory of the Uniform Quantum Computer. Now, one can see the processor hardware agnostic architecture of this compute stack, above the kernel scheduler API.

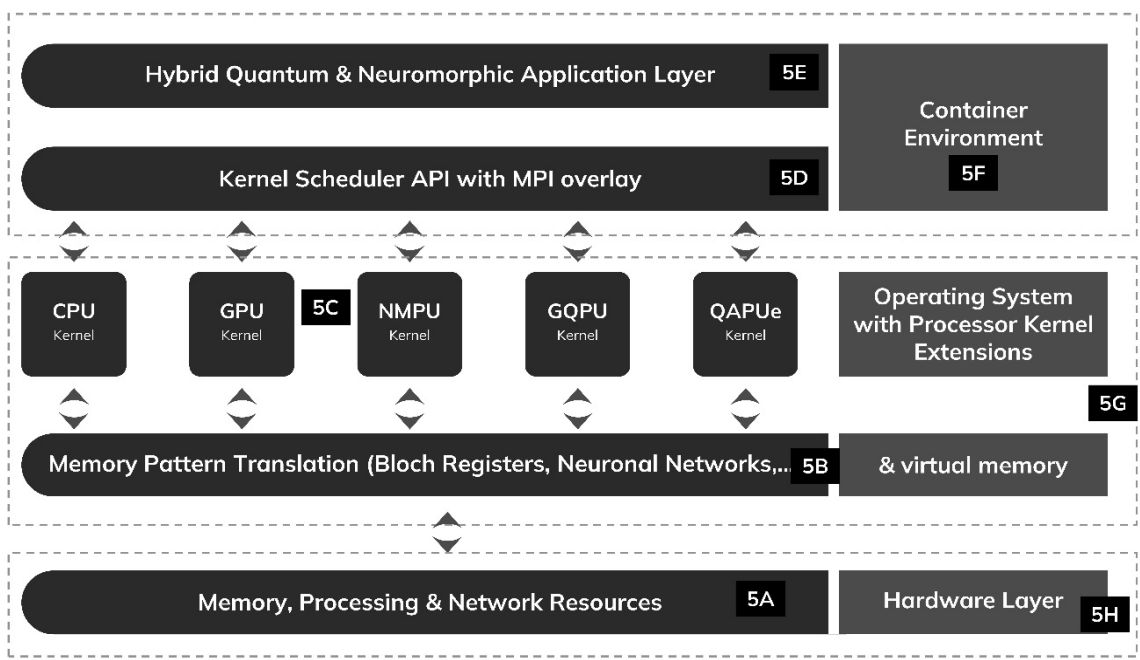

*Figure 5   Uniform Information Processing Stack*

### C. processing the given quantum circuit by transforming all qubits by unitary matrices

In the case of a native quantum processor, this is the task of its arithmetic and logic unit (ALU). Accordingly, the simulated QPU uses the same unit but as a piece of software in the processor kernel extensions of the hybrid quantum computing operating system, carried out by means of the physical resources of a Turing machine, which is part, or the whole of [5A]. Regardless of the physical implementation, the application layer [5E] makes use of the kernel schedulers MPI overlay functionality, which provides the programmer all the useful and available methods of thread parallelization. This makes the Uniform Quantum Computing Model a real high-performance computing environment (HPC) that allows the programmer to distribute applications and also tasks within one application over arbitrary numbers of different processors and compute nodes with the same operating system [5G] running on them. Through its processor kernel extensions and the virtual memory layer, the programmer is able to utilize the full amount of compute resources to one single application and optimize its behavior with the exchange of relevant meta information between the application and the Hybrid Quantum Operating System.

### D. measuring the qubits to retrieve classical information

In the case of the native QPU, this process is physical. The kernel extension of the Hybrid Quantum Operating System is capable of the analogue procedure in a simulated environment, by means of the classical representation of quantum information, as described in chapter III. The memory pattern translation [5B] is able to directly read out the state vector. Thus, this is a feature of Advanced Quantum Inspired Computing, which saves time for certain algorithms, like the Quantum Approximate Optimization Algorithm (QAOA). But if we aim to build a hybrid quantum application, able to run on native and virtual QPUs simultaneously, we have to refrain from such unfair classical advantage, compared to native QPUs. On the other hand, with the parallelization capabilities of the Uniform Quantum Computing Model, we can do better and run the version of the application, using the shortcuts of AQIC automatically on the virtual QPUs and the other version of the same algorithm on the native QPUs simultaneously.

### E. processing classical information

In the Uniform Quantum Computing Model, this feature is evident, since it represents the seamless merger of classical and quantum computing resources on the information theoretical level. But it is worth mentioning that switching back and forth between sequences of classical and quantum information processing of the same compute thread is extremely fast and very close to its theoretical minimum within the Uniform Information Processing Stack, since the data between two such operations should not be moved within the large amount of shared memory between the different instances of the processing units, called the Virtual Processor Instances (VPIs).

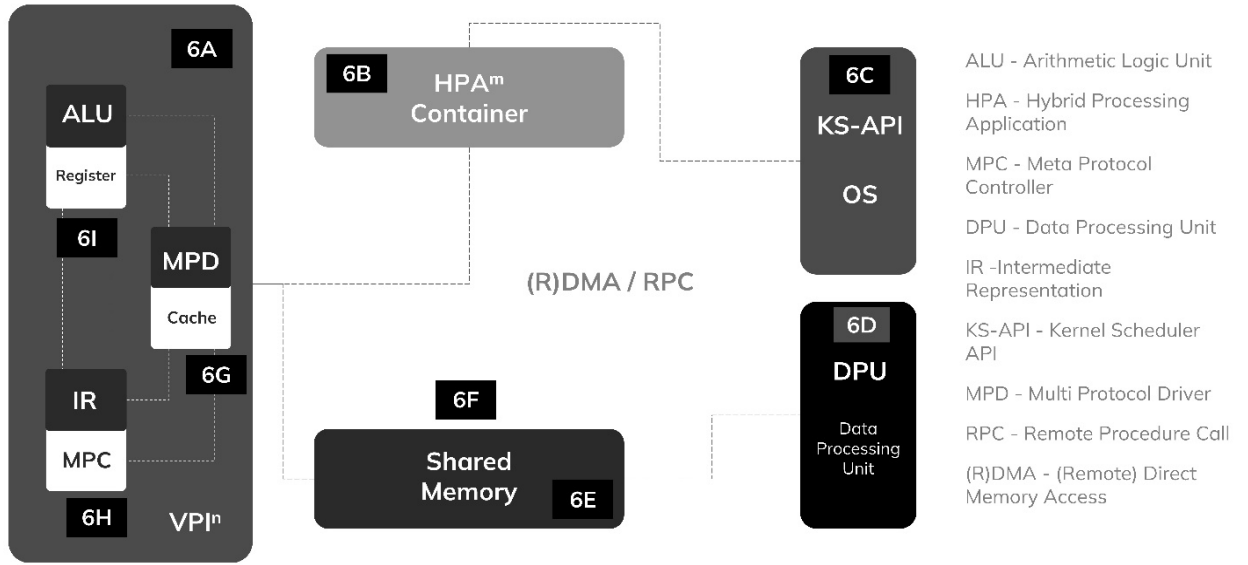

*Figure 6 Virtual Processor Instance*

### VII. VIRTUAL PROCESSOR INSTANCE

A VPI is the processor kernel extension, a memory representation of the structure of a processing unit, irrespective of the physical implementation of such computational appliance. It represents the essence of the functional structure of such a device, represented as software code within a high-performance Turing machine. On the one hand, the VPI should not be seen as an emulation of a physical instance, because it does not reproduce its unwanted, or in the case of quantum processors, erroneous behavior while storing and processing quantum information. In fact, the VPI is an idealization of the device it represents, but with all functional aspects desired from such a device. On the other hand, it is also not a simulator, because it just represents the functional structure of an idealized processing unit, not a fully-fledged computing system, which it is actually part of.

The VPI is integrated in the computing environment as laid out in [6]. Each of these VPIs to the instance n [6A] has access to the full amount of main memory [6E], so the architecture implies shared memory features. The necessary data transfer between the components of the computer is handled with direct memory access (DMA) over a suitable bus system [6F], like a Memory Channel or PCIe. Thus, the architecture joins the group of Non-Uniform Memory Access (NUMA) constructions. Functions are executed on the optimal hardware for their type, favoring the non-moving data strategy of the Uniform Information Processing Model (UIPM) which holds also for instruction code. As a consequence, subroutines are preferably invoked over remote procedure calls (RPC). The VPIs [6A] are coherently constructed, regardless of their physical implementation, and can differ a lot from the idealized structure, represented within the respective Kernel Extensions [5C]. Since each of these elements has direct memory access and is terminated by the same software virtualization layer in the VPI, they are available for high-performance parallelization methods, like the message passing interface standard MPI, used by the Kernel Scheduler API [5D]. This API allows for the applications, which run parallel in the Hybrid Processing Container environment [6B], to parallelize their threads into a universe of Virtual Processing Units [6A], respectively nodes of the whole operating system. Such an operating system with processor kernel extensions and virtual memory [5G] is the first of its kind to span a homogenous and hardware agnostic abstraction (virtualization) layer over heterogenous physical and virtual processing units.

Both the physical and virtual processing units are represented by a functional memory pattern, the VPI [6A], and consist of three major parts as follows:

*Multi Protocol Driver [6G]*

The MPD functions as the driver interface to the operating system [6C], as well as it handling the communication between the VPI as a whole and its two other inner components. It has to be capable of the translation of the different protocols and functions as a switch between the internal components of the VPI and the external systems. The MPD also holds the cache for the virtual processing unit, which is either built by the MPD in memory, if there is no physical implementation of the VPI behind, or it maps the physical cache of a physical (quantum) processing unit into the main memory and thus, provides cache coherency throughout the system, e.g., with a protocol like the Compute Express Link (CXL).

*Meta Protocol Controller [6H]*

The MPC handles the meta information exchanged over the MPD and holds the Intermediate Representation (IR) for the information processing structures, such as quantum circuits – e.g., with the Quantum Assembly Language (QASM), or link patterns for neural networks. This meta information, then, is handling the physical or virtual resources like qubits or neurons.

*Arithmetic & Logic Unit [6I]*

The ALU, as with any processing unit, is the core of the logic and arithmetic operations which are carried out between the registers of the processor. In the case of a gate-based quantum processor, this is a linear algebra representation with matrix operations.

One can see that this newly proposed Uniform Information Processing Model is also very well suited to simulate quantum computers with other hardware, such as matrix processing units (GPUs). In fact, one can explore how to implement algorithms, written for a specific hardware, to a totally different one and what performance impact the result has. The Data Processing Unit [6D] is used to connect many of such memory centric compute nodes to even larger, coherent central memory structures, which can span a whole data center facility with thousands of nodes.

VIII. CONCLUSION

This proposed new computing architecture for high-performance, highly scalable applications in data centers is a turning away from today's execution centered operating systems in our HPC nodes, which differ with any processor type in their singular kernels, toward a memory-centric operating system with kernel extensions for every kind of processing unit, homogenously presented to the application layer and functionally stored in a single, central memory. Today's software development frameworks are fit to support the novel operating system and new libraries and will enable them to take advantage from the hybrid (quantum) computing approach.

The applications built on such a Uniform Information Processing Platform will be able to run on future versions, because the advances in hardware will contribute to the speed-up of the apps but veiled by the intermediate representation of the Virtual Processor Instance. This will not hinder the programmers adding new features based on newer versions of the hardware and operating system – they will then be upwards compatible with future generations of the system. This is a very promising feature for the industry, as it has been with the x86 architecture for more than 40 years, since it saves the investments made in software development, already well above a trillion Euros for the installed base in use today.

In conclusion, we expect to see huge potential from the optimization with computation of hard problems using this novel Uniform Quantum Computing Model, as soon as the new quantum processor technology is ready for exhaustive data center usage, as it allows for Advanced Quantum Inspired High-Performance Computing today, which seamlessly transforms into hybrid quantum computing.